\documentclass[sigconf,screen,authorversion]{acmart}

\AtBeginDocument{%
  \providecommand\BibTeX{{%
    \normalfont B\kern-0.5em{\scshape i\kern-0.25em b}\kern-0.8em\TeX}}}

\setcopyright{acmlicensed}
\copyrightyear{2024}
\acmYear{2024}
\acmBooktitle{Proceedings of the 2024 12th International Conference on Computer and Communications Management (ICCCM '24). Association for Computing Machinery, New York, NY, USA, pp. 93–102, \url{https://doi.org/10.1145/3688268.3688283}}
\acmDOI{}

\usepackage{todonotes}
\usepackage[nolist]{acronym}
\usepackage{siunitx}
\usepackage{tabularx}
\usepackage{array}
\usepackage{caption}
\usepackage{subcaption}
\usepackage{graphicx}
\usepackage{lscape}

\usepackage{tikz}
\newcommand*\circled[1]{\tikz[baseline=(char.base)]{
            \node[shape=circle,draw,inner sep=1pt] (char) {#1};}}

\begin{document}

\title{On the Effect of TSN Forwarding Mechanisms on \\Best-Effort Traffic}

\author{Lisa Maile}
\email{lisa.maile@fau.de}
\orcid{0000-0002-2528-7760}
\affiliation{%
  \institution{Computer Networks and Communication Systems, Friedrich-Alexander-Universität Erlangen-Nürnberg}
  \country{Germany}
}
\authornote{During the publication process, Lisa Maile's affiliation changed to the Institute of Computer and Network Engineering, Technische Universität Braunschweig, Germany.}
\author{Dominik Voitlein}
\affiliation{%
  \institution{Computer Networks and Communication Systems, Friedrich-Alexander-Universität Erlangen-Nürnberg}
  \country{Germany}
}
\author{Anna Arestova}
\affiliation{%
  \institution{Computer Networks and Communication Systems, Friedrich-Alexander-Universität Erlangen-Nürnberg}
  \country{Germany}
}
\author{Abdullah S. Alshra`a}
\affiliation{%
  \institution{Computer Networks and Communication Systems, Friedrich-Alexander-Universität Erlangen-Nürnberg}
  \country{Germany}
}
\author{Kai-Steffen J. Hielscher}
\affiliation{%
  \institution{Computer Networks and Communication Systems, Friedrich-Alexander-Universität Erlangen-Nürnberg}
  \country{Germany}
}
\author{Reinhard German}
\affiliation{%
  \institution{Computer Networks and Communication Systems, Friedrich-Alexander-Universität Erlangen-Nürnberg}
  \country{Germany}
}

\renewcommand{\shortauthors}{Maile and Voitlein et al.}

\begin{abstract}
Time-Sensitive Networking (TSN) enables the transmission of multiple traffic types within a single network. While the performance of high-priority traffic has been extensively studied in recent years, the performance of low-priority traffic varies significantly between different TSN forwarding algorithms. This paper provides an overview of existing TSN forwarding algorithms and discusses their impact on best-effort traffic. The effects are quantified through simulations of synthetic and realistic networks. The considered forwarding mechanisms are Strict Priority (SP), Asynchronous Traffic Shaper (ATS), Credit-Based Shaper (CBS), Enhanced Transmission Selection (ETS), and Time-Aware Shaper (TAS).

The findings indicate that ATS, CBS, and ETS can significantly reduce queuing delays and queue lengths for best-effort traffic when compared to SP and TAS. This effect is enhanced when the reserved bandwidth for high priority queues - using CBS, ATS, or ETS - is reduced to the lowest possible value, within the reserved rate and latency requirements. Specifically, the simulations demonstrate that the choice of forwarding algorithm can improve the performance of low-priority traffic by up to twenty times compared to the least effective algorithm. This study not only provides a comprehensive understanding of the various TSN forwarding algorithms but can also serve as guidance at networks’ design time to improve the performance for all types of traffic.
\end{abstract}

\ccsdesc[500]{Networks~Network design principles}
\ccsdesc[500]{Networks~Packet scheduling}
\ccsdesc[500]{Networks~Network simulations}

\keywords{Time-Sensitive Networking, Low-Priority, Scheduling, Forwarding Algorithms, Simulation}


\maketitle

\section{Introduction}
Many modern communication systems need to integrate different types of traffic for critical and non-critical communications. 
Historically, these systems have relied on separate networks to handle these different requirements. This separation has a significant impact on the cost and weight of the systems - a critical factor for networks such as in automobiles, aircraft, and other vehicles.

To meet the growing demand for large real-time systems that can handle multiple traffic types simultaneously, the IEEE 802.1 Task Group introduced \acf{TSN}~\cite{802Q}. \ac{TSN} is a set of standards and protocols that extend Ethernet for real-time communications. \ac{TSN} is used in a wide range of applications including audio and video broadcasting, industrial automation, automotive communications and aircraft control. A key feature of \ac{TSN} is its ability to integrate traffic of varying criticality within the same network. Hereby, traffic can be divided into hard or soft \textit{real-time} and \textit{non-real-time} traffic~\cite{traffic_criticality}.
Since non-real-time traffic is typically scheduled with the lowest priority, we will use the term \textit{low-priority} and \textit{best-effort} (BE) traffic interchangeably in the following.

The most common forwarding algorithms used in 
\ac{TSN} are \ac{TAS}, \ac{SP}, \ac{ETS}, \ac{CBS}, and \ac{ATS}. The lowest priority queue is typically used for \ac{BE} traffic.
Some forwarding mechanisms have been developed specifically for particular use cases. For example, \ac{TAS} is designed for ultra-low latency traffic, while \ac{CBS} is tailored for audio and video traffic. 
The choice of forwarding algorithm for the highest priority queue has a profound effect on low-priority queues, affecting not only the available bandwidth but also the average delay experienced. Although low-priority traffic does not generally require latency guarantees, it benefits from high data rates and strong average performance.

However, as applications diversify and \ac{QoS} requirements expand, the choice of an appropriate forwarding algorithm becomes increasingly complex.
Moreover, various studies show that - to meet the required guarantees of mixed-type networks - multiple forwarding mechanisms can be applied to high-priority flows~\cite{itans_anna,nasrallah_performance_2019,zhao_quantitative_2022,zhou_simulating_2021}.

This paper investigates the effect of high-priority forwarding mechanisms and their influence on the performance of low-priority queues.
If multiple algorithms can satisfy real-time requirements, the choice can be optimized to take into account the performance of low-priority traffic at design time, thereby improving overall network performance.

To the best of our knowledge, no previous study has systematically compared the impact of different \ac{TSN} forwarding mechanisms on low-priority traffic. Our work provides both theoretical and simulative comparisons to assess the performance of non-real-time traffic, focusing on delay and queue length in both synthetic and automotive network settings. We also investigate the impact of configuration parameters that determine reserved bandwidth per queue on the low priorities. Thus, the results presented in this paper provide insights for the design of future \ac{TSN} networks. 

The paper is organized as follows: Section~\ref{sec:related} reviews and introduces the existing literature. Section~\ref{sec:tsn} introduces the concepts of forwarding mechanisms in TSN, before Section~\ref{sec:effect} describes their theoretical effects on low-priority traffic. Section~\ref{sec:evaluation} then presents various simulation results to quantify these effects. The paper concludes with Section~\ref{sec:conclusion}, which summarizes our findings and contributions.

\section{Related Work}\label{sec:related}
In recent years, many studies have focused on comparing different \ac{TSN} forwarding mechanisms, particularly for high-priority traffic. 

A considerable amount of work has been devoted to analytical latency analysis of \ac{TSN}, as shown by the surveys of Maile et al.~\cite{maile_network_2020} and Deng et al.~\cite{deng_survey_2022}. In this context, Zhao et al.~\cite{zhao_quantitative_2022} conducted an analytical comparison of high-priority traffic, examining latency in networks using \ac{TAS}, \ac{SP}, ATS and CBS, including combinations of these mechanisms. Similarly, Arestova et al.~\cite{itans_anna} focused on an analytical comparison for high-priority traffic SP and TAS networks.

Other studies have used real hardware for their comparisons. Regev et al.~\cite{regev_is_2016} demonstrated initial results in a presentation using hardware switches and proprietary measurement devices to analyze network latency, discussing the effect of traffic bursts in CBS, SP and \ac{WRR}. Bosk et al.~\cite{bosk_simulation_2023} used both simulation and hardware implementations to compare TAS and CBS, with the aim of supporting the practical deployment of TSN in hardware environments.

Using simulation, Geyer et al.~\cite{geyer_evaluation_2013} compared various forwarding mechanisms such as Weighted Fair Queuing (WFQ), Worst-Case Fair Weighted Fair Queuing (WF2Q), First In First Out (FIFO), Deficit Round Robin (DRR), CBS, and SP in an avionics network. Migge et al.~\cite{migge_insights_nodate} compared SP and CBS, and Zhou et al.~\cite{zhou_simulating_2021} additionally compared SP, CBS, ATS, and TAS. TSN has also been compared to other technologies using simulations, such as AFDX~\cite{feng_he_impact_2017}.

However, the above studies focused on comparing the performance of high-priority traffic. To the best of our knowledge, only Geyer et al.~\cite{geyer_evaluation_2013} and Regev et al.~\cite{regev_is_2016} have also quantified the effect on low-priority traffic. Geyer et al. showed that CBS is more beneficial for low-priority traffic than SP. However, they did not observe a significant difference for WFQ and DRR compared to SP due to the specific configurations used in their study. Regev et al. observed that CBS outperformed SP and WRR for low-priority traffic, but the evaluation statements are short. Thus, a full comparison of different forwarding mechanisms in TSN is still missing.

Furthermore, to ensure real-time transmission of high-priority flows, the forwarding parameters must be configured to take into account the latency requirements of high-priority flows. Some works, such as by~\cite{houtan_synthesising_2021,velasco_supporting_2020}, have proposed configuration solutions that aim to minimize the impact of their configurations on low-priority traffic. While the effects are considered in the solutions, these approaches fall short in comparing effects among different forwarding algorithms.

To meet this gap, we provide the first strategic comparison of the impact of all TSN forwarding mechanisms on low-priority traffic.

\section{Overview of TSN Forwarding Algorithms}\label{sec:tsn}

This section introduces the most prominent TSN forwarding algorithms. For each algorithm, the theoretical effects on BE traffic are presented in Section~\ref{sec:effect}. 

\begin{figure}[t]
\begin{center}
\includegraphics[width=\columnwidth]{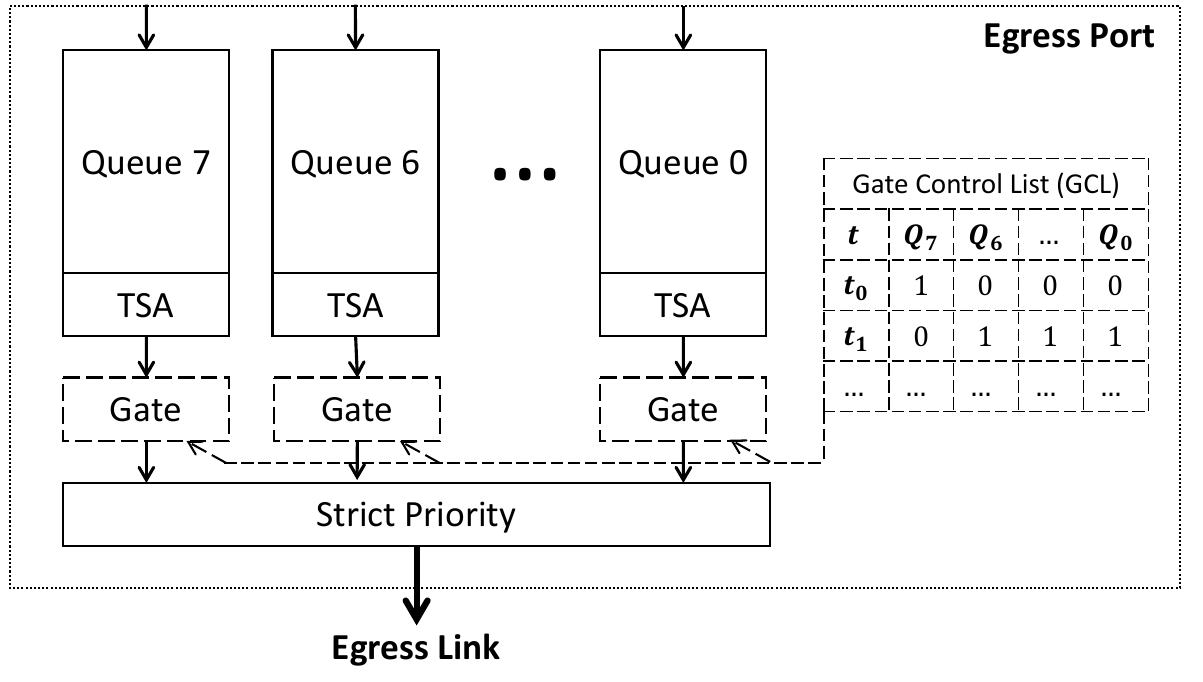}
\end{center}
\caption{TSN forwarding logic per egress port} 
\label{fig:egressport}
\end{figure}

In \ac{TSN}, each egress port shares a physical link for all streams, and the selected forwarding algorithms determine the transmission order for frames in each port. We will use the terms stream and flow interchangeably in the following.
Figure~\ref{fig:egressport} shows the logic of packet forwarding in \ac{TSN} egress ports. The combination of these features will be referred to as the \textit{forwarding algorithm}. 

First, frames are placed into \textbf{queues} according to their priority, so multiple flows can share the same queue. \ac{TSN} allows up to eight queues. For the purposes of this analysis it is assumed, without loss of generality, that each priority is assigned to a single queue. In scenarios where multiple priorities are grouped within a single queue, the performance of priorities in the same queue is identical, as there is no differentiation in their treatment during the data forwarding process.

As shown in Fig.\ref{fig:egressport}, a \textbf{\ac{TSA}} determines whether the frames in a queue are eligible for transmission. In addition, queues can have optional \ac{TT} \textbf{gates} which control the transmission times of frames, i.e., when the gates are open. Finally, if multiple queues have eligible frames and open gates, the transmission order is determined by \textbf{\ac{SP}}, meaning that higher priority frames (with higher queue IDs) are sent first. 

We define $\mathcal{H}$ as the set of successive queues $q_i\in \mathcal{H}$ with $0 < i \leq 7$ serving high-priority traffic. We assume that at least one queue 
serves BE traffic with the lowest priority. For our analysis, all queues in $\mathcal{H}$ have the same \ac{TSA}. Unless otherwise stated, BE queues are assumed to be forwarded without \ac{TSA}, i.e., served in order of priority.

\subsection{Strict Priority}
\ac{TSN} allows \ac{SP} forwarding when all gates are open and no \ac{TSA} is selected.
In \ac{SP}, frames with the highest priority are always transmitted first~\cite{802Q}. 
This ensures that the most important frames are forwarded as soon as possible. 

\subsection{Scheduled Traffic}
\ac{TT} gates allow traffic in TSN to be separated in time, thus requiring global clock synchronization in the network~\cite{AS}. This approach prioritizes traffic by assigning specific transmission time slots to different queues. This concept is formally described in the TSN standards as ``scheduled traffic''~\cite{802Q}.

Scheduled traffic mechanisms divide time into cycles of equal length. Within these cycles, sub-time slots are allocated for the transmission of real-time data packets. Gate Control Lists (GCLs) define the behaviour of the gates, and the state of each gate is either open (1) or closed (0). This is illustrated on the right side of Fig.~\ref{fig:egressport}, where, for example, during $[t_0,t_1)$ only queue $q_7$ is open. 
In addition, guard bands are placed in front of each time slot of critical traffic to prevent new transmissions. This ensures that critical traffic never has to wait for other transmissions. 

One implementation of the scheduled traffic concept that has received considerable research attention is called \ac{TAS}.
In TAS systems, both end stations and switches are configured to synchronize the transmission of frames. Most approaches of TAS follow a no-wait procedure, i.e., flows are scheduled in the network at times when they are guaranteed not to interfere with other traffic~\cite{menth_survey_tas_2023}. To accomplish this, each flow is given an offset for transmission. Also, each time a frame arrives at a switch, the schedule ensures that the TT gate is configured to be open and that no other flows are waiting. No further \ac{TSA} is applied in the queues. Thus, TAS effectively minimizes queuing delay for high-priority traffic.

An alternative approach using the scheduled traffic concept is called \ac{CQF}. \ac{CQF} operates using two queues, opening them alternatively with equal time intervals. Newly arriving real-time frames are placed into the currently closed queue and are transmitted in the next interval. High-priority flows are assigned to the intervals, and the sum of the transmission time of all frames assigned to one interval has to be shorter than the interval duration itself. Consequently, the end-to-end latency for a flow is dependent only on the interval duration and the number of hops on the path. 
Within each interval, lower priority traffic is transmitted after the high-priority traffic with \ac{SP}, using the leftover time in the interval.

\subsection{Credit-Based Shaper}
CBS~\cite{802Q} was introduced in 2009 as the first implementation for a \ac{TSA}. It was specifically designed to reduce the negative impact of high priorities on BE traffic. This is achieved by introducing idle times during high-priority transmissions. The queue parameter \textit{idleSlope} defines the minimum guaranteed bandwidth for each queue, but also imposes an upper limit on the queue's transmission, or in other words, each queue receives its allocated bandwidth, but no more. 

\begin{figure}[t]
\begin{center}
\includegraphics[width=\columnwidth]{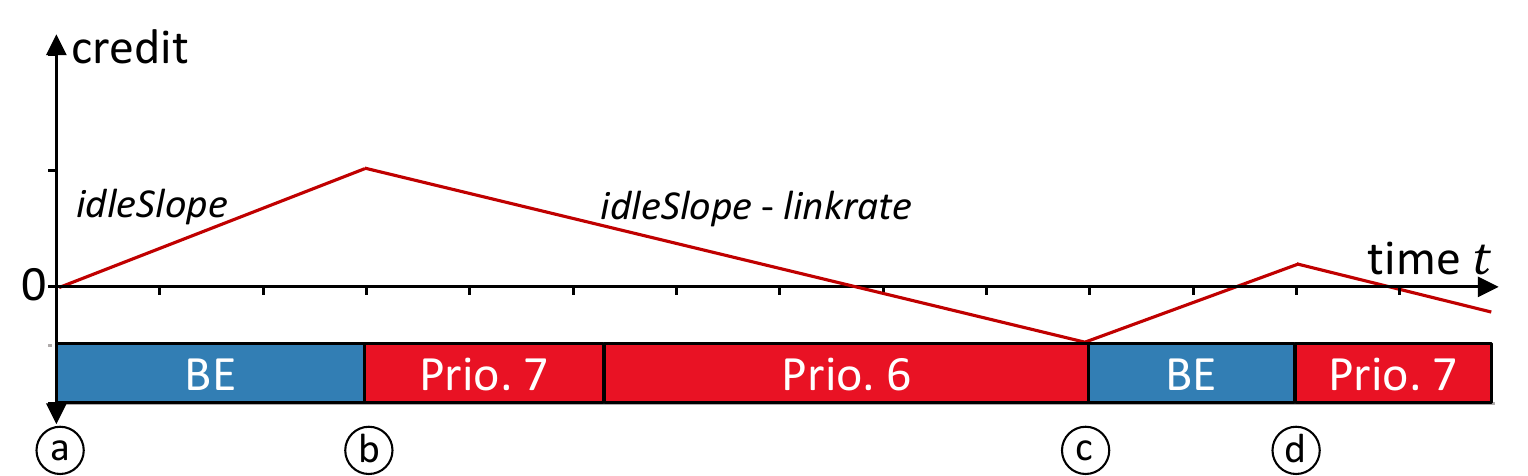}
\end{center}
\caption{Credit evolution for CBS forwarding} 
\label{fig:cbs}
\end{figure}

CBS assigns a credit to each queue, which determines whether frames are eligible for transmission: only if the credit is positive ($\ge 0$), frames can start their transmission.
Figure~\ref{fig:cbs} illustrates transmission for CBS. If the transmission of other priority queues delays an eligible transmission of a queue (Fig.~\ref{fig:cbs}, phase \circled{a} and \circled{c}), the credit of the queue increases with the rate \textit{idleSlope}. Each transmission decreases the credit of the queue with the rate \textit{idleSlope-linkrate} (Fig.~\ref{fig:cbs}, phase \circled{b} and \circled{d}). If a queue is empty and the credit is positive, the credit is reset to zero.
As frames are not allowed to begin their transmission while the credit is negative, this allows other frames, such as BE, to be transmitted at that time.

\subsection{Asynchronous Traffic Shaper}
\ac{ATS}~\cite{802Q} was designed to prevent burstiness cascading within the network, thereby facilitating latency analysis of flows and improving the performance of lower priorities.

In its original proposal, ATS implemented two levels of queues, called shaped and shared queues. Frames were first stored in the shaped queues to wait for re-shaping. After re-shaping, they are placed in the shared queues, which are the eight priority queues of TSN. This algorithm has been adapted to the TSN standards by the introduction of eligibility times: Each flow has an individual state that determines the time at which the frame becomes eligible for transmission. Both implementations have been proven to be identical~\cite{boyer_equivalence_ats}.

The re-shaping algorithm of ATS follows a token-bucket approach, called the Token Bucket Emulation (TBE) algorithm. For each flow, the bucket size is defined by the \textit{committed burst size} (cbs) parameter and the token generation rate is defined by the \textit{committed information rate} (cir) parameter. 
As a result, ATS introduces re-shaping per flow, unlike CBS, which introduces re-shaping per queue.

\subsection{Enhanced Transmission Selection}
The TSN mechanism \ac{ETS}~\cite{802Q} defines general requirements for \acp{TSA}.
It defines that each queue is allocated a configurable percentage of the total bandwidth. While the standard does not specify a particular algorithm, it does suggest the use of a round-robin approach. We will assume \ac{DRR}~\cite{shreedhar_drr_definition_1995} in the following, as it takes into account the size of the packets and thus offers more fairness between flows.

The \ac{DRR} algorithm adds a deficit counter to each queue. It works in a cyclic manner, traversing the queues sequentially. When the round-robin mechanism reaches a queue that is not empty, it increases the queue's deficit counter by a preset \textit{quantum}.
After a frame has been transmitted, the deficit counter is reduced by the number of bytes transmitted. A queue can continue transmission until the remaining deficit is insufficient for the next packet. The remaining deficit is stored for the next cycle, and the DRR algorithm moves on to the next non-empty queue. 

\section{Effect on Lower Priorities} \label{sec:effect}
The forwarding algorithms described in Section~\ref{sec:tsn} affect the performance of low-priority queues in different ways. This section provides an overview of the theoretical differences between the algorithms. 

\begin{figure}[t]
\begin{center}
\includegraphics[width=1\columnwidth]{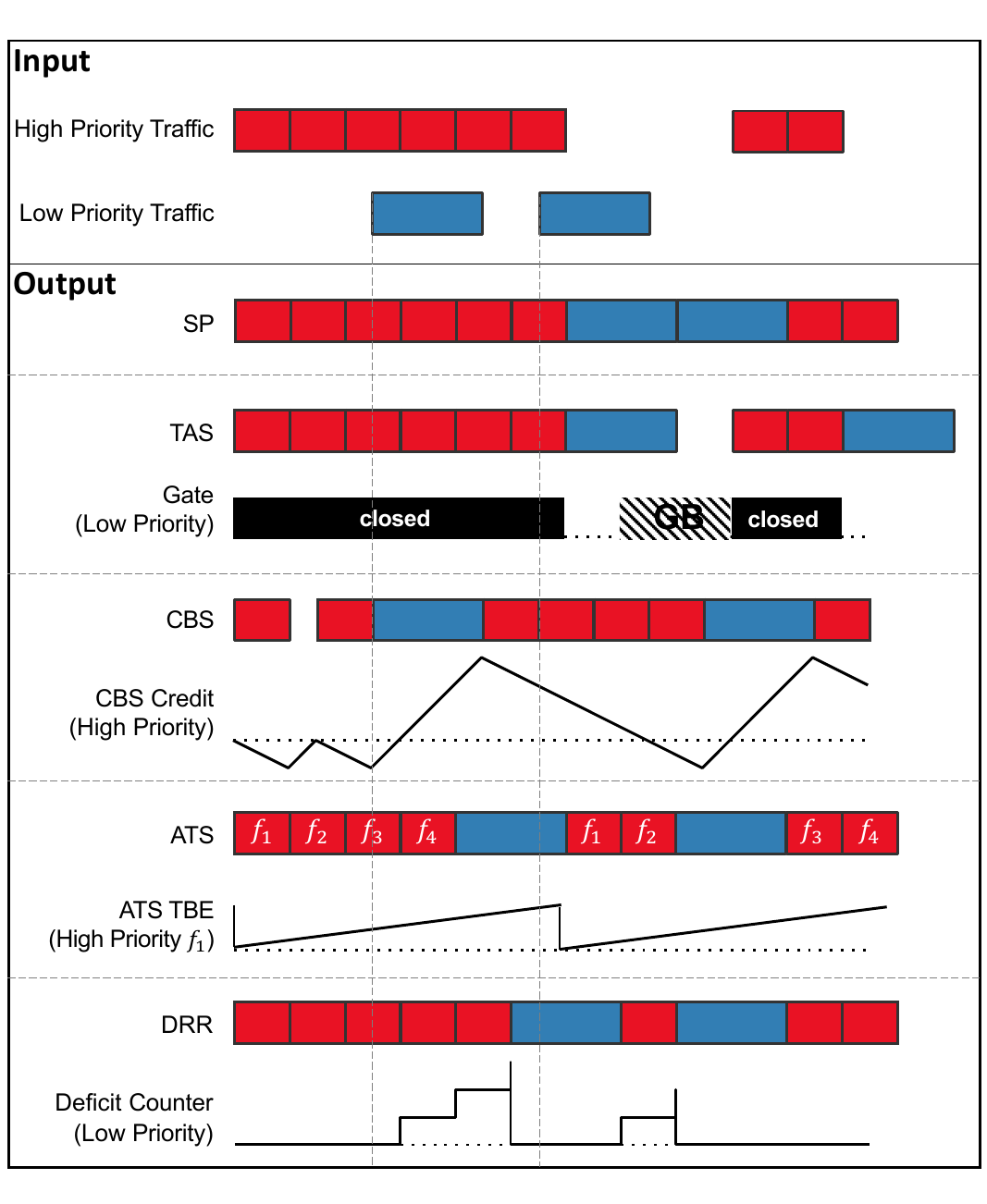}
\end{center}
\caption{Different forwarding algorithms compared} 
\label{fig:theory_effect}
\end{figure}

Figure~\ref{fig:theory_effect} illustrates example inputs of high- and low-priority traffic to compare the main differences between the handling in TSN forwarding algorithms. Note that the effects are for illustration only, the figure does not allow conclusions on overall performance of the forwarding algorithms. This will be quantified in Section~\ref{sec:evaluation}.

\subsubsection*{Strict Priority}
SP has been studied extensively and is known to delay low-priority frames significantly because SP forwarding is not fair - meaning that the queues do not share bandwidth. Whenever high-priority frames are queued, they are given priority. This results in significant delays for low-priority frames, a problem also known as ``starvation''.
This can be seen in Fig.~\ref{fig:theory_effect}, where with SP the burst of high-priority frames is fully transmitted before the low-priority traffic is served.

\subsubsection*{Scheduled Traffic}
\begin{figure}[t]
\begin{center}
\includegraphics[width=\columnwidth]{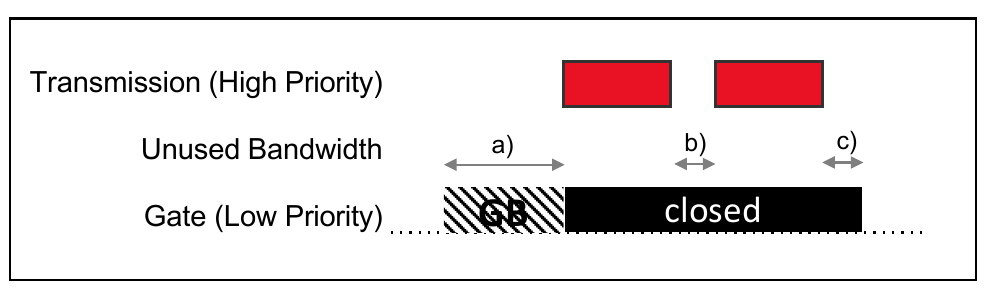}
\end{center}
\caption{Different effects that reduce the effective bandwidth when using TAS} 
\label{fig:TAS_lost}
\end{figure}

Solutions with time-separation of traffic, as implemented in TAS, define the impact on low-priority queues at design time.
Figure~\ref{fig:TAS_lost} illustrates the critical aspect of TAS.

The first effect, labelled a) in Fig.~\ref{fig:TAS_lost}, is due to the implementation of guard bands (GB) before the open period of high-priority queues. Guard bands can reduce the bandwidth available to low-priority frames if these frames arrive during an active guard band period, consequently being delayed in transmission.
Some TAS configuration strategies have been developed to reduce the number of guard bands, e.g., by minimizing the number of critical transmission slots~\cite{anna_arestova_journal_2023,jingates}. However, these countermeasures often lead to combined and extended intervals of high-priority transmission, which are not fully utilized. Thus, it has a trade-off since such an extended transmission time for high-priority queues notably increases the average delay, jitter, and burstiness of low-priority traffic. 

The second effect of TAS, illustrated by b), results from the strictly timed transmission intervals for high-priority frames. Since high-priority frames are allocated explicit transmission times, it often happens that the interval between two high-priority frames is insufficient to transmit or open the gate for other frames. This further increases the waiting time for low-priority frames.

Finally, TAS requires that all high-priority frames are guaranteed to transmit during their assigned gate open times. Therefore, if there is jitter in the transmission time of frames, this jitter must be accounted for and the gate open times must be extended. In realistic scenarios, this jitter is inevitable due to hardware delays and clock inaccuracy, and increases with the length of the path. For example, in Fig.~\ref{fig:TAS_lost}, the second packet could be delayed by c) on its path, so the gates must also cover this delay.

In summary, the performance of low-priority traffic in TAS is strictly dependent on the transmission timing of high-priority frames and the configuration of the GCLs in the network. As seen in Fig.~\ref{fig:theory_effect}, TAS degrades the performance of low-priority traffic even more than SP due to the overheads described.

Finally, \ac{CQF} operates on a different mechanism, where frames of higher priorities are queued during one interval and then transmitted in the subsequent interval with the highest priority. Within each interval, CQF employs SP scheduling.
However, since CQF stores the frames for one interval, they are explicitly transmitted as a burst in the next interval. Consequently, low-priority packets are transmitted only after periodic bulks of high-priority data.
The duration of these bulks depends on the choice of the interval time, as this defines the time that high-priority traffic is queued. Thus, lower interval times improve the performance of low-priority traffic. However, as flows are assigned to the intervals, higher interval times allow for more high-priority flows, but increase their end-to-end latency. When interval times approach zero, CQF performance for low-priority traffic approaches that of SP by definition, with performance degrading for higher interval times.
The interval times are entirely defined by the rate and latency requirements of high-priority flows. As these parameters are not considered in this work, quantitative analysis and comparison of CQF is not possible within the scope of this paper. 

\subsubsection*{Credit-Based Shaper}
The main idea of CBS forwarding is the introduction of credits per queue, which effectively creates time slots available for low-priority traffic. Lower priority queues are guaranteed to receive all remaining bandwidth, which is \textit{linkrate}$-$ $\sum_{\forall i \in \mathcal{H}} \textit{idleSlope}_{q_i}$.

The resulting effects can be seen in Fig.~\ref{fig:theory_effect}. First, CBS is non-work-conserving~\cite{
goos_analysis_non-work-conserving_1995}, i.e., it delays packets even if no other packet is transmitted, as seen after the first high-priority transmission in the figure. This shaping effect on high-priority traffic also affects subsequent network hops, as high-priority traffic arrives less bursty.

Second, only if high-priority queues have positive credit, low-priority traffic will be delayed. 
Consequently, low-priority traffic can be transmitted even if bursts of high-priority frames arrive, as long as $\sum_{\forall i \in \mathcal{H}} \textit{idleSlope}_{q_i}$ $ \le \textit{linkrate}$.

\subsubsection*{Asynchronous Traffic Shaping}
Both CBS and ATS delay high-priority frames to allow low-priority traffic to be transmitted. The primary difference between ATS and CBS is the method of traffic shaping: ATS shapes traffic on a per-flow basis, as opposed to the per-class shaping used by CBS.
When multiple different flows arrive at a network node, ATS allows frames from different flows to be transmitted in sequence as a burst. Only frames from the same flow are shaped. Figure~\ref{fig:theory_effect} illustrates this difference by assuming that the arriving packets belong to different flows $f_i,i\in\{1,2,3,4\}$. Each flow then has its own token bucket shaper.
Again, ATS is non-work conserving, as it delays packets regardless of the presence of low-priority traffic.

Thus, ATS can allow higher bursts per queue than CBS, but no per flow burst. 
Both mechanisms smooth the traffic, which is beneficial in terms of delay at the next hop. However, the quantification of this difference for lower priorities has been missing before and will be evaluated in the next section.

\subsubsection*{Enhanced Transmission Selection}
For ATS and CBS, BE queues have no \ac{TSA} implemented and are still guaranteed to receive residual bandwidth. However, unlike ATS and CBS, DRR is work conserving, meaning queues are allowed to transmit at full link capacity if no other traffic is present. Therefore, if low-priority queues do not implement DRR as \ac{TSA}, the DRR queues would behave exactly like SP for the low-priority queues. Instead, low-priority queues could also implement DRR and be assigned the remaining quantum value to ensure that they do not suffer from starvation. As otherwise no difference to SP could be observed, we will assume in the following that BE queues also implement DRR.
 
Unlike CBS and ATS, DRR queues cannot start transmitting as soon as a packet is received and no high-priority queue is available, but must accumulate quantum at packet reception. This introduces additional delays for each packet, as shown for the first BE packet in Fig.~\ref{fig:theory_effect}. 

Additionally, as DRR is work conserving, 
traffic can be transmitted in bursts without being shaped, leading to increased network delay at subsequent nodes. However, introducing times when no packet is sent, as seen for CBS after the first packet, can lead to increased overall delay at that hop as transmission is shifted. This leads to the situation where DRR can send the second BE packet earlier than CBS in Fig.~\ref{fig:theory_effect}.

\begin{figure*}
\centering
\begin{subfigure}{.49\textwidth}
    \centering
    \includegraphics[width=0.65\linewidth]{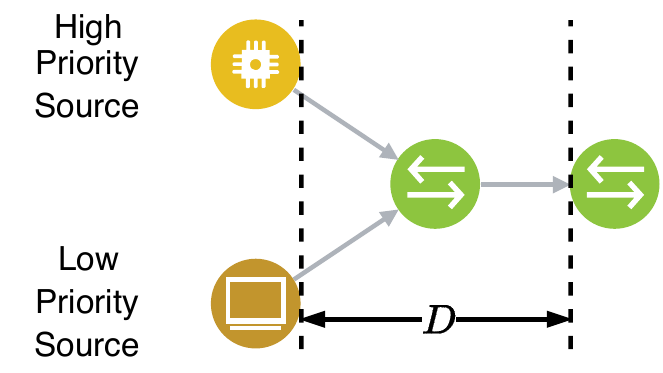}
    \caption{One hop network}
    \label{fig:oneHopNetwork}
\end{subfigure}
\begin{subfigure}{.49\textwidth}
    \centering
    \includegraphics[width=0.65\linewidth]{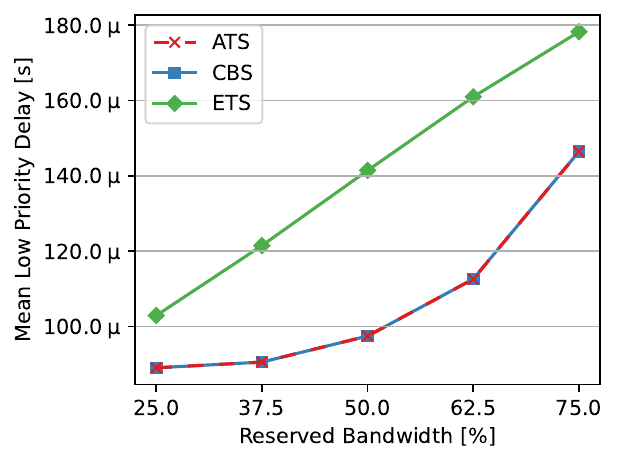}
    \caption{Delay with high-priority flow using 25\% of linkrate}
    \label{fig:overprovisioning}
\end{subfigure}%
\caption{Evaluations on different configuration in one-hop network}
\label{fig:one-hop}
\end{figure*}

\section{Evaluation}\label{sec:evaluation}
This section quantifies the impact of the effects described in the previous sections. We use simulations to measure the delay and backlog of \ac{ATS}, \ac{CBS}, \ac{ETS}, \ac{SP}, and \ac{TAS}. To determine the \ac{GCL} for \ac{TAS} we used~\cite{itans_anna}, an algorithm based on the well-known no-wait scheduling approach~\cite{no-wait_mascis_2022}. We assume no hardware jitter and perfect clock synchronization, so that the effect of TAS on lower priorities is minimized.

In the following simulations, high-priority traffic is assigned priority 7, while BE is assigned priority 6. The links operate at \SI{100}{\mega\bit\per\second} bandwidth with no preemption.
The configurations for all simulations are provided in Tables~\ref{tab:part1} and~\ref{tab:part2}.
In some figures, the tables introduce variables that denote the x-axis of the figure, which are $R$ for the reserved bandwidth, $U_h$ for high priority utilization, and $b_h$ and $b_l$ for burst length of high priority~(HP) and low priority~(LP) traffic, respectively. Traffic originates either from deterministic sources (Det.) with periodic, equal-length bursts, or from \ac{MMPP} sources with two states (fast and slow).
We use the same traffic characteristics and load for all forwarding mechanisms to get a fair comparison for the low-priority delays, sending one flow per source node.

For each forwarding mechanism in the following evaluation, the \ac{QoS} provided to high-priority traffic is different. Configuring a network to meet delay requirements is non-trivial and its own area of research. A comparison between the algorithms when considering equal high-priority deadlines, as well as a comparison of other scheduling approaches for \ac{TAS}, is left for future work.

In the following, our plots focus on mean delay values, as the variability in delay values is large due to the randomness in inter-arrival times, packet sizes, and the interactions between low- and high-priority packets. The variability of delay is illustrated through boxplots at the end of this section.

\subsection{One Hop} \label{sec:oneHop}
Figure~\ref{fig:oneHopNetwork} shows the first setup with a single hop. Latency measurements are taken from source transmission to sink reception.
The allocated bandwidth for high-priority queues, defined by the \textit{committed information rate} in ATS, \textit{idleSlope} in CBS, and \textit{quantum} in ETS, directly influences the delay. At minimum, the reserved bandwidth must match the long-term transmission rates of queues. Increasing the reserved bandwidth for a queue reduces its transmission delay, as shown, e.g., in~\cite{maile_journal_2022}. So more bandwidth for high-priority traffic improves its delay performance, but reduces the residual bandwidth for low-priority traffic, thus, increasing its latency.

Figure~\ref{fig:overprovisioning} examines the average delay for low-priority traffic when the reserved bandwidth for the high-priority flow is overprovisioned (1 to 3 times). All forwarding mechanisms show increased delays with overprovisioning, with DRR having higher average delays, but ATS and CBS being more sensitive to overprovisioning. Careful configuration is therefore essential to balance high-priority and low-priority performance to meet all real-time deadlines without exceeding the required rate.

\begin{figure*}
\centering
\begin{subfigure}{.49\textwidth}
    \centering
    \includegraphics[width=0.65\linewidth]{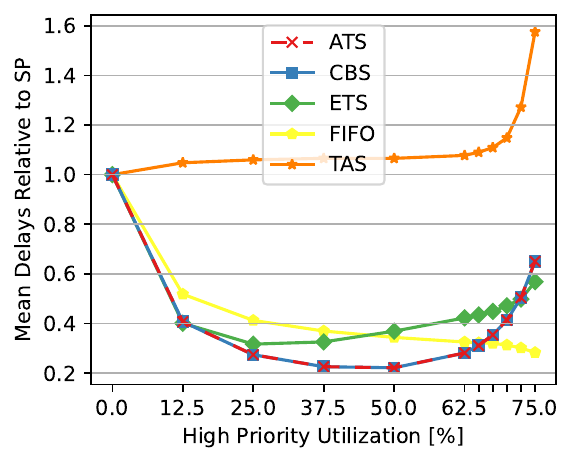}
    \caption{One hop network delays relative to \ac{SP}}
    \label{fig:oneHopRelDelay}
\end{subfigure}%
\begin{subfigure}{.49\textwidth}
    \centering
    \includegraphics[width=0.65\linewidth]{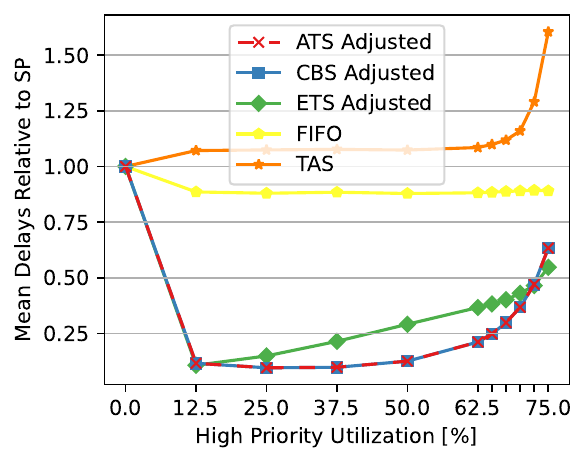}
    \caption{One hop network backlog relative to \ac{SP}}
    \label{fig:oneHopRelBacklog}
\end{subfigure}
\caption{One hop network delays and backlog relative to \ac{SP}}
\label{fig:oneHop}
\end{figure*}

In the following, we assume that the reserved bandwidth parameters for all high-priority traffic match the long-term rate.
To show the differences between the forwarding mechanisms, Fig.~\ref{fig:oneHop} compares all measurements to SP. The mean delay and backlog increases for increasing high-priority utilization. \ac{TAS} performs worse than SP because of the effects described in Section~\ref{sec:effect}. \ac{ATS}, \ac{CBS} and \ac{ETS} have up to four times lower delays and backlogs than SP. This demonstrates the effect of these algorithms in allowing low-priority packets to pass between packets of a high-priority burst. It can be seen that ETS performs superior to ATS and CBS for higher utilizations of high priority traffic. This is due to the fact that ETS does not prioritize high-priority traffic but cyclically iterates through all queues. 
For reference, Fig.~\ref{fig:oneHop} also shows the results with \ac{FIFO} scheduling with only one priority for all traffic. However, \ac{FIFO} does not allow any packet prioritization as required for real-time transmission. Therefore, it is not considered in the following setups.

\begin{figure*}
\centering
\begin{subfigure}{.49\textwidth}
    \centering
    \includegraphics[width=0.65\linewidth]{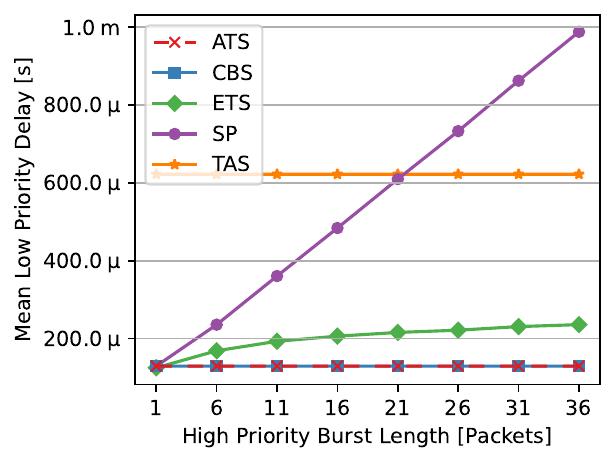}
    \caption{Delays for varying high-priority burst length}
    \label{fig:burstLengthHP}
\end{subfigure}
\begin{subfigure}{.49\textwidth}
    \centering
    \includegraphics[width=0.65\linewidth]{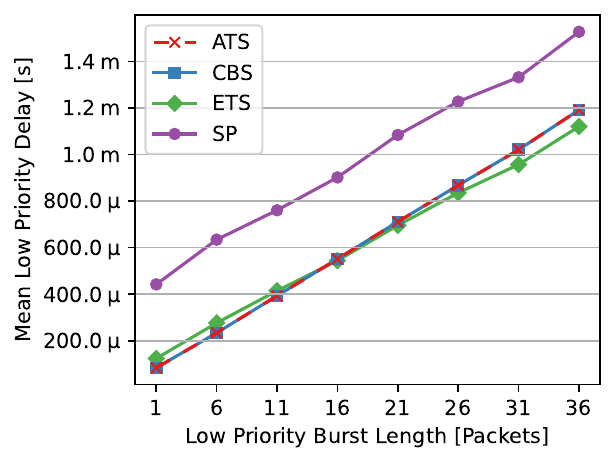}
    \caption{Delays for varying low-priority burst length}
    \label{fig:burstLengthLP}
\end{subfigure}%
\caption{Impact evaluation of bursts on the delay at 50\% network utilization}
\label{fig:burstLength}
\end{figure*}

To show the effect of traffic burstiness, Fig.~\ref{fig:burstLength} measures the delay for different burst sizes of high- and low-priority traffic.
Figure~\ref{fig:burstLengthHP} shows that the number of high-priority packets per burst has a significant impact on delays. For \ac{TAS} the delay depends only on the \ac{GCL} which remained static throughout the experiment. For SP, the increase in delay is directly proportional to the burst length. All other forwarding mechanisms distribute the packets evenly over time. Therefore, they have a similar performance regardless of the burst length. \ac{ETS} introduces more delay than \ac{ATS} and \ac{CBS}, but significantly less than SP. 
Figure~\ref{fig:burstLengthLP} shows the effect of changing the burstiness of the low-priority source. 
The low priority keeps its utilization constant at 20\%, but sends packets deterministically in bursts of varying lengths. Longer bursts result in increasing delays for all forwarding mechanisms. Note that the relative differences between the forwarding mechanisms decrease. For example, with a burst length of one, the delay of SP is about four times that of \ac{CBS}. At 36, it is only 1.2 times higher. In general, to improve the performance of BE traffic, less bursty transmission should be preferred.

\subsection{Star Topology}
\label{sec:starTopology}

\begin{figure*}
\centering
\begin{subfigure}{.49\textwidth}
    \centering
    \includegraphics[width=0.55\linewidth]{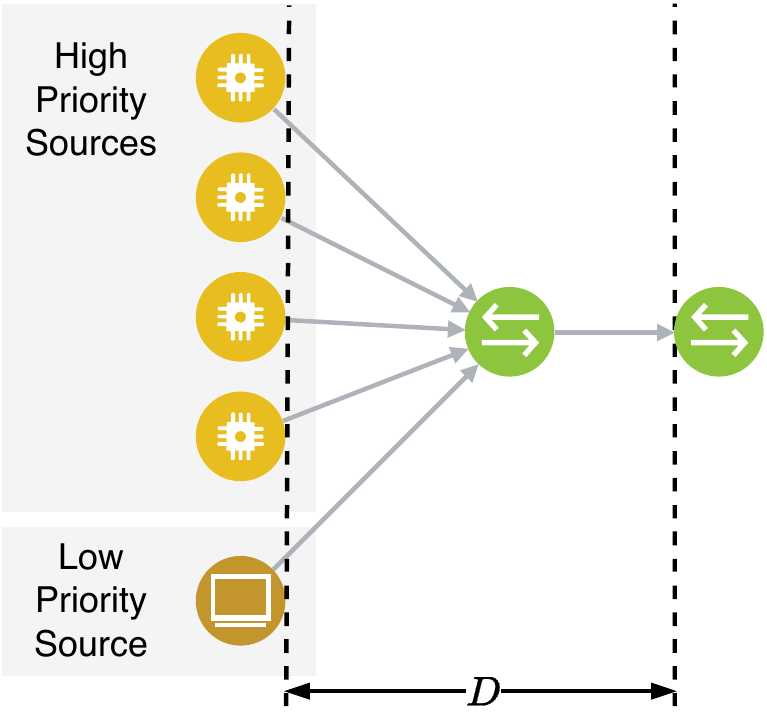}
    \caption{Star network}
    \label{fig:starNetwork}
\end{subfigure}
\begin{subfigure}{.49\textwidth}
    \centering
    \includegraphics[width=0.65\linewidth]{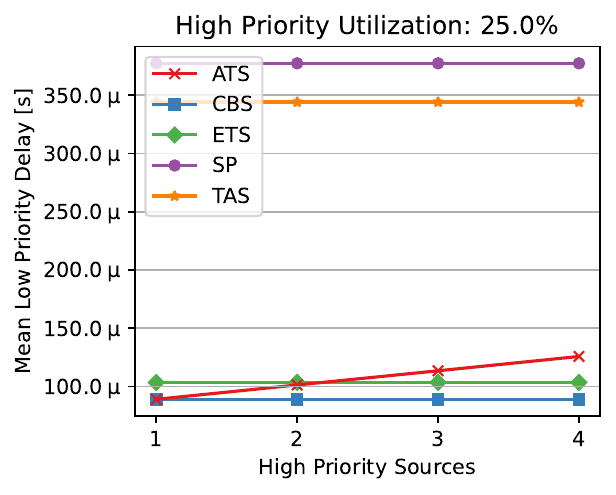}
    \caption{Star network delay at 25\% utilization}
\end{subfigure}
\begin{subfigure}{.49\textwidth}
    \centering
    \includegraphics[width=0.65\linewidth]{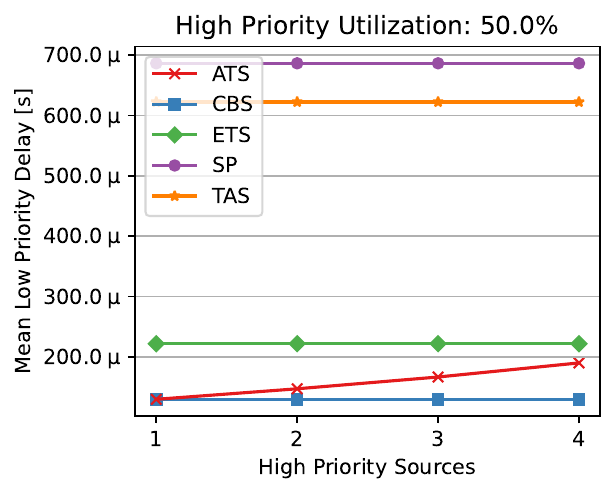}
        \caption{Star network delay at 50\% utilization}
\end{subfigure}
\begin{subfigure}{.49\textwidth}
    \centering
    \includegraphics[width=0.65\linewidth]{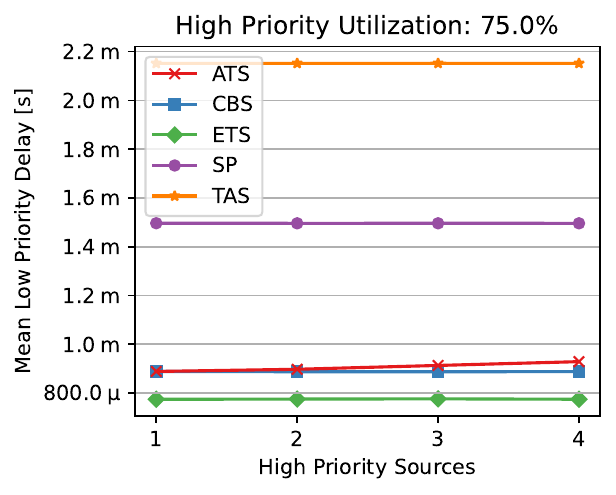}
        \caption{Star network delay at 75\% utilization}
\end{subfigure}
\caption{Delays in a star topology for 1-4 sources}
\label{fig:starDelay}
\end{figure*}

Next, we analyze how the number of flows affects the delays. We use the network in Fig.~\ref{fig:starNetwork} with one flow per source and the total load for the high-priority traffic split evenly between one to four sources. Figure~\ref{fig:starDelay} shows the results for different network utilizations. Only the delays for ATS change as the number of streams increases because it is the only algorithm that shapes each flow individually. Otherwise, the results are similar to the single-hop network. It can be seen that for ATS increasing the number of distinct flows reduces the performance of low-priorities, as the flows can then be transmitted in bursts, as described in Section~\ref{sec:effect}.

\subsection{Tree Topology}
\label{sec:treeTopology}

\begin{figure*}
\centering
\begin{subfigure}{.49\textwidth}
    \centering
    \includegraphics[width=0.4\linewidth]{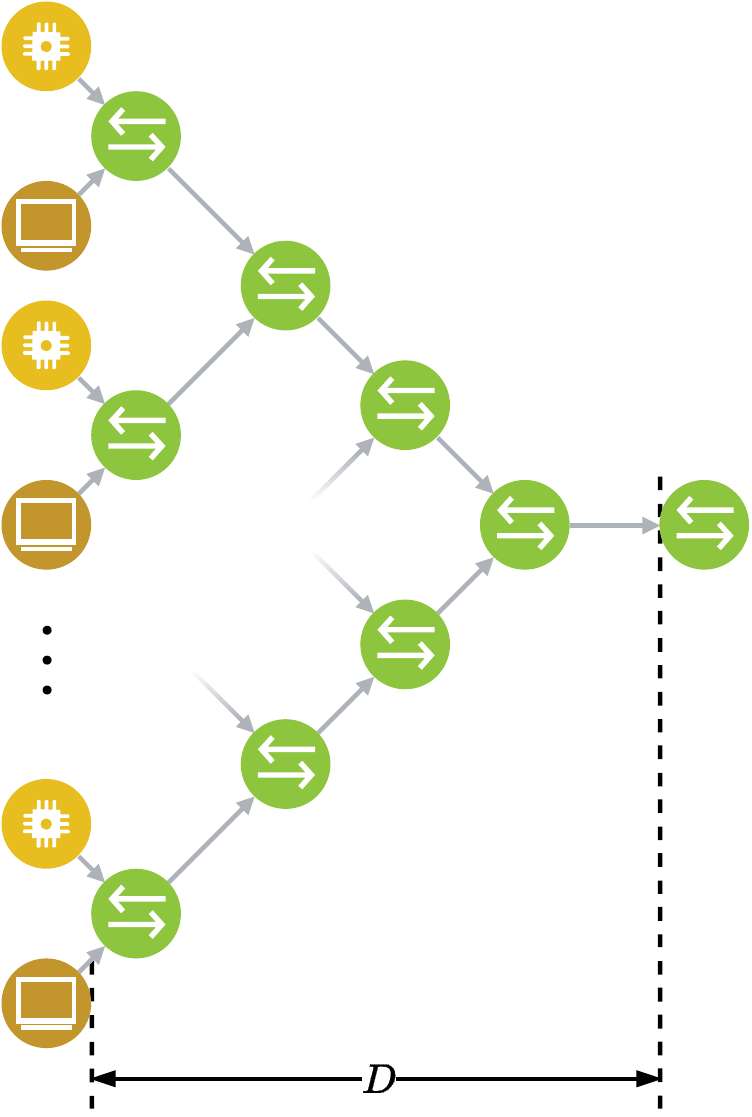}
    \label{fig:treeNetwork}
\end{subfigure}
\begin{subfigure}{.49\textwidth}
    \centering
    \includegraphics[width=0.7\linewidth]{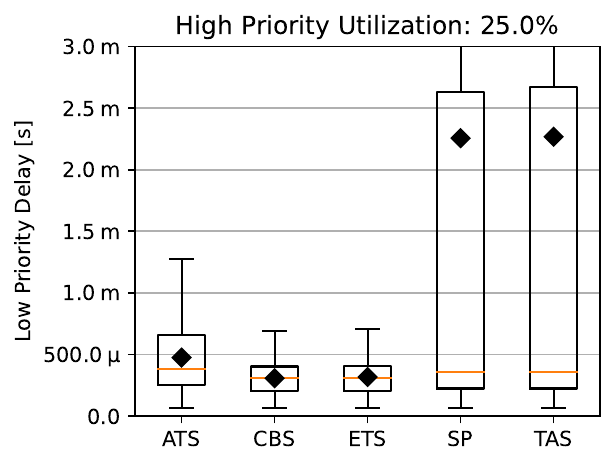}
\end{subfigure}
\begin{subfigure}{.49\textwidth}
    \centering
    \includegraphics[width=0.7\linewidth]{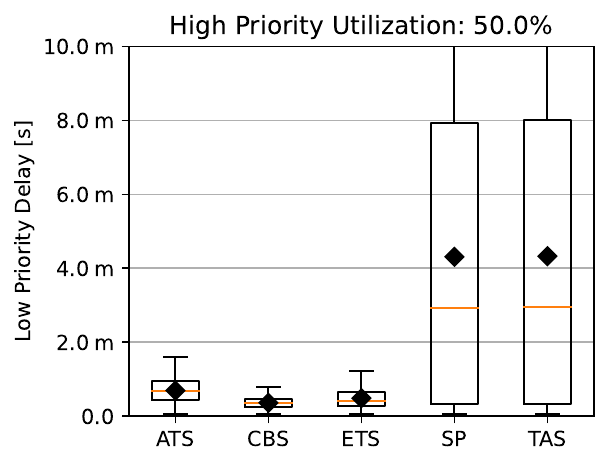}
\end{subfigure}
\begin{subfigure}{.49\textwidth}
    \centering
    \includegraphics[width=0.7\linewidth]{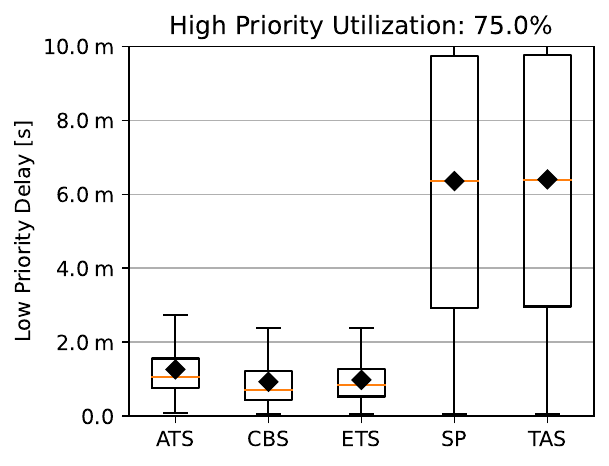}
\end{subfigure}
\caption{Delays in a tree topology}
\label{fig:tree}
\end{figure*}

The last synthetic evaluation focuses on the tree topology shown in Fig.~\ref{fig:tree}. It has eight low and eight high-priority sources sending to the same destination. Again, the load 
is divided equally between all sources. 

The results in Fig.~\ref{fig:tree} for different utilization levels show again that SP and TAS have a strong impact on low-priority traffic. However, due to the increased multiplexing and the higher number of flows, \ac{ATS} has higher latencies than \ac{CBS} and ETS, which was not the case before.

\subsection{Automotive Network}
\label{sec:car}

\begin{figure*}
\centering
\begin{subfigure}{.5\textwidth}
    \centering
    \includegraphics[width=0.65\linewidth]{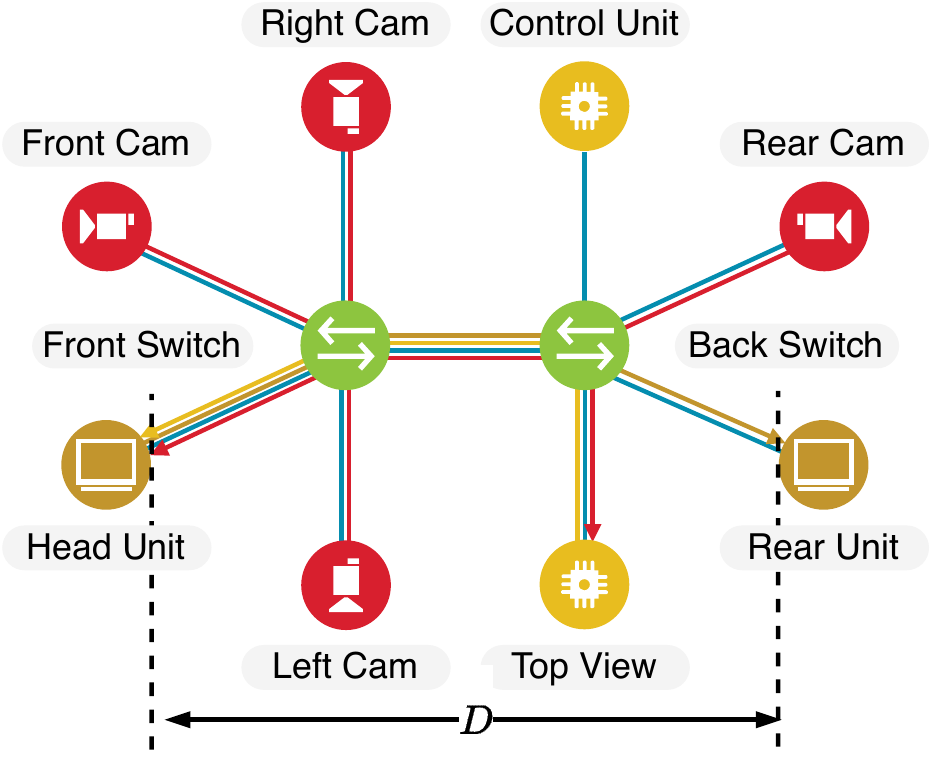}
    \caption{Automotive driver-assistance and multimedia network~\cite{CarExampleQueck}}
    \label{fig:carNetwork}
\end{subfigure}%
\begin{subfigure}{.5\textwidth}
    \centering
    \includegraphics[width=0.65\linewidth]{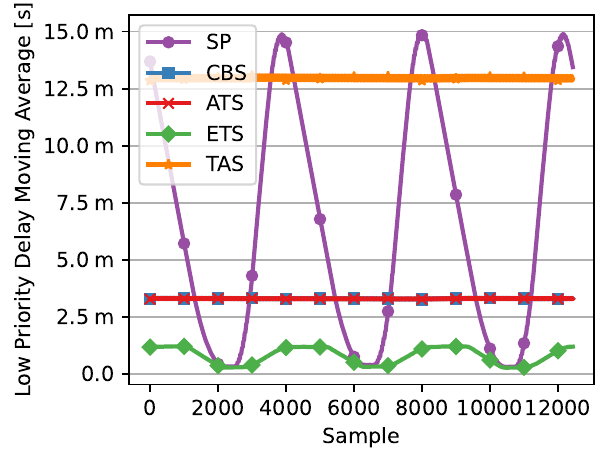}
    \caption{Delays over time with a 1000 sample moving average}
    \label{fig:carBurstOverTime}
\end{subfigure}
\caption{Tests on automotive network}
\label{fig:automotive}
\end{figure*}

\begin{figure*}
\centering
\begin{subfigure}{.5\textwidth}
    \centering
    \includegraphics[width=0.7\linewidth]{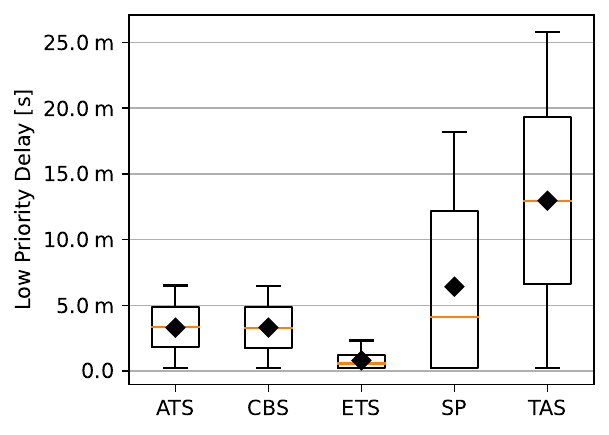}
    \caption{Delays with 43 packet low-priority bursts}
    \label{fig:carBurstDelay}
\end{subfigure}%
\begin{subfigure}{.5\textwidth}
    \centering
    \includegraphics[width=0.7\linewidth]{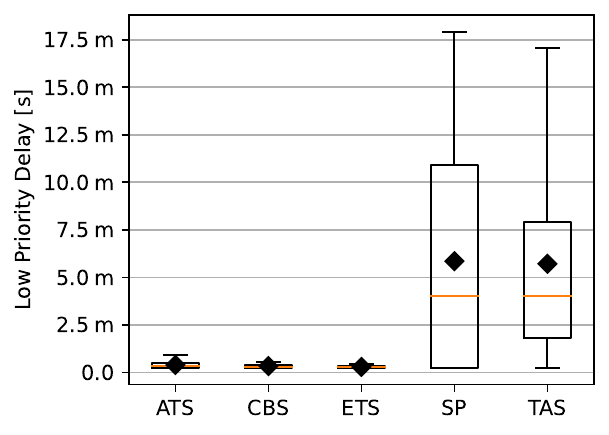}
    \caption{Exponentially distributed low-priority inter-arrival times}
    \label{fig:carExpDelay}
\end{subfigure}
\caption{Delay evaluation in automotive network}
\label{fig:automotive_delay}
\end{figure*}

The last simulation evaluates an automotive driver assistance and multimedia network to quantify the effects in a realistic scenario. The topology is shown in Fig.~\ref{fig:carNetwork} and the traffic characteristics are taken from~\cite{CarExampleQueck}. 

This network shows an interesting behavior because of the periodic traffic transmission. All streams have defined sending intervals with no randomness.  Therefore, the utilization of the network significantly varies over time. Figure~\ref{fig:carBurstOverTime} shows the resulting fluctuations of the latency. When both camera and multimedia bursts occur simultaneously, the delay for SP forwarding is high (e.g., around samples 0, 4000, and 8000). Conversely, around samples 2000 and 6000, the multimedia packets arrive between the camera bursts. Therefore, the latency is much lower, resulting in high jitter when using SP. In contrast, neither \ac{ATS} nor \ac{CBS} result in notable jitter for BE traffic because they distribute the high-priority frames evenly over time. ETS shows jitter, but to a much lesser extent than SP. The shaping effect introduced by the non-work conserving idle times in ATS and CBS and the prioritization of high priority traffic comes at the cost of higher average delays when compared to ETS.

This can also be observed in Fig.~\ref{fig:carBurstDelay}, which shows the delays for BE in the network. The differences are similar to the results of the synthetic evaluation, but not as extreme. As already seen in Fig.~\ref{fig:burstLengthLP} the differences between the forwarding algorithms decrease with a higher low-priority burstiness. In the automotive network, the burst length is $43\text{  packets} \cdot \SI{1542}{\byte}$. With this large burst, the delay values are closer together than in other experiments. Also, \ac{TAS} performs considerably worse. In this case, due to the more complex network topology and transmission periods, the schedule opens and closes the gates frequently. Therefore, the guard bands take up more of the time. In addition, the low-priority gates do not always stay open long enough to transmit a packet, further delaying BE packets.

Finally, we repeat the simulation with a Markovian low-priority source instead of sending the data periodically to reduce the burstiness.
Figure~\ref{fig:carExpDelay} shows the results. The latency of SP stayed at about \SI{6}{\milli\second} because frames had to wait until the high-priority bursts are finished. All other forwarding mechanisms improved immensely with \ac{ATS}, \ac{CBS} and \ac{ETS} at around \SI{300}{\micro\second}. This reduction in burstiness therefore improves the performance of ATS, ETS and CBS, resulting in their average performance being 20 times better than \ac{SP}.

\section{Conclusion and Future Work}\label{sec:conclusion}

This paper has investigated the effects of different Time-Sensitive Networking (TSN) forwarding algorithms on best-effort traffic, contributing to the understanding of TSN's ability to support multiple traffic types. Our simulations, both synthetic and realistic, show that Asynchronous Traffic Shaper (ATS), Credit-Based Shaper (CBS), and Enhanced Transmission Selection (ETS) are more efficient in reducing delays and queue lengths for low-priority traffic compared to Strict Priority (SP) and Time-Aware Shaper (TAS), especially when their forwarding parameters are minimized to match the transmission rates. In particular, the choice of forwarding mechanisms can immensely impact low-priority traffic, with up to twenty times better performance than the least effective forwarding mechanism.

In the future, we plan to investigate different high-priority configuration solutions that take into account the delay requirements of high-priority traffic, and quantify their impact on low-priority traffic. In addition, TSN allows for the combination of forwarding mechanisms, such as the combination of TAS with CBS. Thus, a more detailed investigation of these combinations on best-effort traffic could be a valuable direction for future research. 

In summary, the presented results can offer practical guidance for the selection of forwarding mechanism in future TSN network to improve the network performance across all traffic types.

\begin{acronym}
\acro{TSN}[TSN]{Time Sensitive Networking}
\acro{SP}[SP]{Strict Priority}
\acro{BE}[BE]{Best-Effort}
\acro{TSA}[TSA]{Transmission Selection Algorithm}
\acro{TT}[TT]{Time-Triggered}
\acro{CBS}[CBS]{Credit-Based Shaper}
\acro{ATS}[ATS]{Asynchronous Traffic Shaper}
\acro{ETS}[ETS]{Enhanced Transmission Selection}
\acro{DRR}[DRR]{Deficit Round Robin}
\acro{WRR}[WRR]{Weighted Round Robin}
\acro{TAS}[TAS]{Time-Aware Shaper}
\acro{MMPP}[MMPP]{Markov-Modulated Poisson Process}
\acro{FIFO}[FIFO]{First In - First Out}
\acro{GCL}[GCL]{Gate Control List}
\acro{CQF}[CQF]{Cyclic Queuing and Forwarding}
\acro{QoS}[QoS]{Quality of Service}
\end{acronym}

\begin{table*}[]
\centering
\footnotesize
\begin{tabular}{|l|cc|cc|cc|cc|}
\hline
                                          & \multicolumn{2}{c|}{\textbf{Fig. 5}}                                                                                                                                                      & \multicolumn{2}{c|}{\textbf{Fig. 6}}                                                                                                                                                       & \multicolumn{2}{c|}{\textbf{Fig. 7a}}                                                                                                                                                  & \multicolumn{2}{c|}{\textbf{Fig. 7b}}                                                                                                                                                      \\ \hline
\textbf{Priority}                         & \multicolumn{1}{c|}{High}                                                                    & Low                                                                                        & \multicolumn{1}{c|}{High}                                                                     & Low                                                                                        & \multicolumn{1}{c|}{High}                                                                 & Low                                                                                        & \multicolumn{1}{c|}{High}                                                                     & Low                                                                                        \\ \hline
\textbf{Source Type}                      & \multicolumn{1}{c|}{Det.}                                                                    & MMPP                                                                                       & \multicolumn{1}{c|}{Det.}                                                                     & MMPP                                                                                       & \multicolumn{1}{c|}{Det.}                                                                 & MMPP                                                                                       & \multicolumn{1}{c|}{MMPP}                                                                     & Det.                                                                                       \\ \hline
\textbf{Trans. Matrix {[}1/s{]}}          & \multicolumn{1}{c|}{}                                                                        & \begin{tabular}[c]{@{}c@{}}→fast:   10\\      →slow: 40\end{tabular}                       & \multicolumn{1}{c|}{}                                                                         & \begin{tabular}[c]{@{}c@{}}→fast:   10\\      →slow: 40\end{tabular}                       & \multicolumn{1}{c|}{}                                                                     & \begin{tabular}[c]{@{}c@{}}→fast:   10\\      →slow: 40\end{tabular}                       & \multicolumn{1}{c|}{\begin{tabular}[c]{@{}c@{}}→fast:   10\\      →slow: 40\end{tabular}}     &                                                                                            \\ \hline
\textbf{Inter-Arrival Rates {[}1/s{]}}          & \multicolumn{1}{c|}{}                                                                        & \begin{tabular}[c]{@{}c@{}}fast:   3750\\      slow: 2690\end{tabular}                     & \multicolumn{1}{c|}{}                                                                         & \begin{tabular}[c]{@{}c@{}}fast:   3750\\      slow: 2690\end{tabular}                     & \multicolumn{1}{c|}{}                                                                     & \begin{tabular}[c]{@{}c@{}}fast:   3750\\      slow: 2690\end{tabular}                     & \multicolumn{1}{c|}{\begin{tabular}[c]{@{}c@{}}fast: 7500\\      slow: 5380\end{tabular}}     &                                                                                            \\ \hline
\textbf{Avg. Rate / Flow {[}Mbit/s{]}}  & \multicolumn{1}{c|}{25}                                                                      & 20                                                                                         & \multicolumn{1}{c|}{100*$U_h$}                                                                 & 20                                                                                         & \multicolumn{1}{c|}{50}                                                                   & 20                                                                                         & \multicolumn{1}{c|}{50}                                                                       & 20                                                                                         \\ \hline
\textbf{Packet Sizes {[}byte{]}}          & \multicolumn{1}{c|}{1000}                                                                    & $\mathcal{U}(200, 1400)$ & \multicolumn{1}{c|}{1000}                                                                     & $\mathcal{U}(200, 1400)$ & \multicolumn{1}{c|}{1000}                                                                 & $\mathcal{U}(200, 1400)$ & \multicolumn{1}{c|}{1000}                                                                     & $\mathcal{U}(200, 1400)$ \\ \hline
\textbf{Burst Length {[}packets{]}}       & \multicolumn{1}{c|}{20}                                                                      &                                                                                            & \multicolumn{1}{c|}{20}                                                                       &                                                                                            &  \multicolumn{1}{c|}{$b_h$}                                                               &                                                                                            & \multicolumn{1}{c|}{}                                                                         & $b_l$                                                                                       \\ \hline
\textbf{ATS Param. {[}bit{]}{[}Mbit/s{]}} & \multicolumn{1}{c|}{\begin{tabular}[c]{@{}c@{}}cbs:   8000\\      cir: R\end{tabular}}       &                                                                                            & \multicolumn{1}{c|}{\begin{tabular}[c]{@{}c@{}}cbs:   8000\\      cir: 100*$U_h$\end{tabular}} &                                                                                            & \multicolumn{1}{c|}{\begin{tabular}[c]{@{}c@{}}cbs:   8000\\      cir: 5000\end{tabular}} &                                                                                            & \multicolumn{1}{c|}{\begin{tabular}[c]{@{}c@{}}cbs:   8000\\      cir: 100*$U_h$\end{tabular}} &                                                                                            \\ \hline
\textbf{CBS IdleSlope {[}\%{]}}           & \multicolumn{1}{c|}{\begin{tabular}[c]{@{}c@{}}Fig.   5a) R\\      Fig. 5b) 25\end{tabular}} &                                                                                            & \multicolumn{1}{c|}{$U_h$}                                                                     &                                                                                            & \multicolumn{1}{c|}{50}                                                                   &                                                                                            & \multicolumn{1}{c|}{$U_h$}                                                                     &                                                                                            \\ \hline
\textbf{ETS Quantum {[}byte{]}}           & \multicolumn{1}{c|}{250}                                                                     & 250                                                                                        & \multicolumn{1}{c|}{1000*$U_h$}                                                                & 250                                                                                        & \multicolumn{1}{c|}{50000}                                                                & 250                                                                                        & \multicolumn{1}{c|}{1000*$U_h$}                                                                & 250                                                                                        \\ \hline
\textbf{Topology}                         & \multicolumn{2}{c|}{One Hop}                                                                                                                                                              & \multicolumn{2}{c|}{One Hop}                                                                                                                                                               & \multicolumn{2}{c|}{One Hop}                                                                                                                                                           & \multicolumn{2}{c|}{One Hop}                                                                                                                                                               \\ \hline
\end{tabular}%
\caption{Simulation Configurations (Part 1)}
\label{tab:part1}
\end{table*}

\begin{table*}[]
\centering
\footnotesize
\begin{tabular}{|l|cc|cc|ccc|}
\hline
                                          & \multicolumn{2}{c|}{\textbf{Fig. 8}}                                                                                                                                                         & \multicolumn{2}{c|}{\textbf{Fig. 9}}                                                                                                                                                         & \multicolumn{3}{c|}{\textbf{Fig. 10}}                                                                                                                                                                                                                             \\ \hline
\textbf{Priority}                         & \multicolumn{1}{c|}{High}                                                                       & Low                                                                                        & \multicolumn{1}{c|}{High}                                                                       & Low                                                                                        & \multicolumn{1}{c|}{High}                                                                   & \multicolumn{1}{c|}{Mid}                                                                    & Low                                                                   \\ \hline
\textbf{Source Type}                      & \multicolumn{1}{c|}{Det.}                                                                       & MMPP                                                                                       & \multicolumn{1}{c|}{Det.}                                                                       & MMPP                                                                                       & \multicolumn{1}{c|}{Det.}                                                                   & \multicolumn{1}{c|}{Det.}                                                                   & Det.                                                                  \\ \hline
\textbf{Trans. Matrix {[}1/s{]}}          & \multicolumn{1}{c|}{}                                                                           & \begin{tabular}[c]{@{}c@{}}→fast:   10\\      →slow: 40\end{tabular}                       & \multicolumn{1}{c|}{}                                                                           & \begin{tabular}[c]{@{}c@{}}→fast:   10\\      →slow: 40\end{tabular}                       & \multicolumn{1}{c|}{}                                                                       & \multicolumn{1}{c|}{}                                                                       &                                                                       \\ \hline
\textbf{Inter-Arrival Rates {[}1/s{]}}          & \multicolumn{1}{c|}{}                                                                           & \begin{tabular}[c]{@{}c@{}}fast: 3750\\      slow: 2690\end{tabular}                       & \multicolumn{1}{c|}{}                                                                           & \begin{tabular}[c]{@{}c@{}}fast: 468.8\\      slow: 336.3\end{tabular}                     & \multicolumn{1}{c|}{}                                                                       & \multicolumn{1}{c|}{}                                                                       &                                                                       \\ \hline
\textbf{Avg. Rate / Flow {[}Mbit/s{]}}  & \multicolumn{1}{c|}{100*$U_h$/N}                                                                 & 20                                                                                         & \multicolumn{1}{c|}{100*$U_h$/8}                                                                 & 2.5                                                                                        & \multicolumn{1}{c|}{0.672}                                                                  & \multicolumn{1}{c|}{17.19}                                                                  & 15.91                                                                 \\ \hline
\textbf{Packet Sizes {[}byte{]}}          & \multicolumn{1}{c|}{1000}                                                                       & $\mathcal{U}(200, 1400)$ & \multicolumn{1}{c|}{1000}                                                                       & $\mathcal{U}(200, 1400)$ & \multicolumn{1}{c|}{84}                                                                     & \multicolumn{1}{c|}{1542}                                                                   & 1542                                                                  \\ \hline
\textbf{Burst Length {[}packets{]}}       & \multicolumn{1}{c|}{24/N}                                                                       &                                                                                            & \multicolumn{1}{c|}{20}                                                                         &                                                                                            & \multicolumn{1}{c|}{10}                                                                     & \multicolumn{1}{c|}{46}                                                                     & 43                                                                    \\ \hline
\textbf{ATS Param. {[}bit{]}{[}Mbit/s{]}} & \multicolumn{1}{c|}{\begin{tabular}[c]{@{}c@{}}cbs:   8000\\      cir: 100*$U_h$/N\end{tabular}} &                                                                                            & \multicolumn{1}{c|}{\begin{tabular}[c]{@{}c@{}}cbs:   8000\\      cir: 100*$U_h$/8\end{tabular}} &                                                                                            & \multicolumn{1}{c|}{\begin{tabular}[c]{@{}c@{}}cbs:   672\\      cir: 0.672\end{tabular}}   & \multicolumn{1}{c|}{\begin{tabular}[c]{@{}c@{}}cbs:   12336\\      cir: 17.19\end{tabular}} &                                                                       \\ \hline
\textbf{CBS IdleSlope {[}\%{]}}           & \multicolumn{1}{c|}{$U_h$}                                                                       &                                                                                            & \multicolumn{1}{c|}{$U_h$*F/8}                                                                   &                                                                                            & \multicolumn{1}{c|}{\begin{tabular}[c]{@{}c@{}}front:   2.7\\      back: 4.71\end{tabular}} & \multicolumn{1}{c|}{front:   0.516}                                                         &                                                                       \\ \hline
\textbf{ETS Quantum {[}byte{]}}           & \multicolumn{1}{c|}{1000*$U_h$}                                                                  & 250                                                                                        & \multicolumn{1}{c|}{1000*$U_h$*F/8}                                                              & 250                                                                                        & \multicolumn{1}{c|}{\begin{tabular}[c]{@{}c@{}}front:   27\\      back: 47.1\end{tabular}}  & \multicolumn{1}{c|}{front:   516}                                                           & \begin{tabular}[c]{@{}c@{}}front:   250\\      back: 250\end{tabular} \\ \hline
\textbf{Topology}                         & \multicolumn{2}{c|}{Star with N+1 Sources}                                                                                                                                                   & \multicolumn{2}{c|}{Tree, 8LP + 8HP, F flows at switch}                                                                                                         & \multicolumn{3}{c|}{Automotive}                                                                                                                                                                                                                                   \\ \hline
\end{tabular}%
\caption{Simulation Configurations (Part 2)}
\label{tab:part2}
\end{table*}


\bibliographystyle{ACM-Reference-Format}
\bibliography{references}

\end{document}